\documentclass[12pt]{article}
\usepackage{amsmath,amssymb}
\usepackage[dvips]{graphicx}
\usepackage{cite}

\newcommand{\half}{\frac{1}{2}}
\newcommand{\nn}{\nonumber}

\newcommand{\morder}[1]{{\cal O}\left(#1 \right)}

\newcommand{\eq}[1]{(\ref{eq:#1})}

\newcommand{\re}{{\rm Re\, }}
\newcommand{\pp}{||}

\newcommand{\bbibitem}[1]{\bibitem{#1}\marginpar{#1}}

\def\Label#1{\label{#1}%
  \smash{\hbox to0pt{\raise1ex\hbox{\tiny[#1]}\hss}}}
\def\noLabels{\let\Label=\label}
\def\nobbibitem{\let\bbibitem=\bibitem}

\newcommand{\be}{\begin{equation}}
\newcommand{\ee}{\end{equation}}
\newcommand{\bea}{\begin{eqnarray}}
\newcommand{\eea}{\end{eqnarray}}

\newcommand{\beq} {\begin{equation}}
\newcommand{\eeq} {\end{equation}}
\newcommand{\beqa} {\begin{eqnarray}}
\newcommand{\eeqa} {\end{eqnarray}}

\newcommand{\mC}{\mathcal{C}}

\newcommand{\mE}{\mathcal{E}}

\newcommand{\vev}[1]{\left\langle#1\right\rangle}

\def\vek{\vec{k}}

\def\ie{\emph{i.e.}}
\def\eg{\emph{e.g.}}
\def\la{\lambda}

\def\om{\omega}

\newcommand{\inv}[1]{\frac{1}{#1}}

\begin{document}

\begin{flushright}
{\it CP$^\textrm{\it 3}$-Origins: 2009-21 } \\
HIP-2009-28/TH
\end{flushright}

\vskip 2cm \centerline{\Large {\bf Electrostatics approach to closed string }}
 \vskip 2mm
\centerline{\Large {\bf pair production from a decaying D-brane}}
 \vskip 1cm
\renewcommand{\thefootnote}{\fnsymbol{footnote}}
\centerline{{\bf Niko
Jokela,$^{1,2}$\footnote{najokela@physics.technion.ac.il} Matti
J\"arvinen,$^{3}$\footnote{mjarvine@cp3.sdu.dk} and Esko
Keski-Vakkuri$^{4}$\footnote{esko.keski-vakkuri@helsinki.fi}}}
\vskip .5cm \centerline{\it ${}^{1}$Department of Physics}
\centerline{\it Technion, Haifa 3200, Israel}
\centerline{\it
${}^{2}$Department of Mathematics and Physics}
\centerline{\it University of Haifa at Oranim, Tivon 36006, Israel}
%\centerline{\it Tivon 36006, Israel}
\centerline{\it ${}^{3}$CP$^\textrm{\it 3}$-Origins,} %University of Southern Denmark}
\centerline{\it Campusvej 55, DK-5230 Odense M, Denmark}
\centerline{\it
${}^{4}$Helsinki Institute of Physics }
\centerline{\it P.O.Box 64, FIN-00014 University of
Helsinki, Finland}

\setcounter{footnote}{0}
\renewcommand{\thefootnote}{\arabic{footnote}}

\begin{abstract}
We consider the emission of two closed string tachyons from a decaying D$p$-brane in bosonic string theory. We
study the high energy limit of the emission amplitude by using an analogue to electrostatics.
The case where the emitted strings have equal energies is analyzed in detail,
and we obtain a relatively simple result for the emission amplitude. We identify expected poles for s- and t-channel emission. In the high energy limit, the amplitude decays exponentially ($\sim e^{-2\pi \omega}$) or faster in the energies $\omega$ of the closed strings.
\end{abstract}

\newpage

%%%%%%%%%%%%%%%%%%%%%%%%%%%%%%%%%%%%%%%%%%%%%%%%%%%%%%%%%%%%%%%%%%%%%%%%%%%%%%%%%%%%%%%%%%%%%%%%%%%%%%
%%%%%%%%%%%%%%%%%%%%%%%%%%%%%%%%%%%%%%%%%%%%%%%%%%%%%%%%%%%%%%%%%%%%%%%%%%%%%%%%%%%%%%%%%%%%%%%%%%%%%%
%%%%%%%%%%%%%%%%%%%%%%%%%%%%%%%%%%%%%%%%%%%%%%%%%%%%%%%%%%%%%%%%%%%%%%%%%%%%%%%%%%%%%%%%%%%%%%%%%%%%%%
%%%%%%%%%%%%%%%%%%%%%%%%%%%%%%%%%%%%%%%%%%%%%%%%%%%%%%%%%%%%%%%%%%%%%%%%%%%%%%%%%%%%%%%%%%%%%%%%%%%%%%
%%%%%%%%%%%%%%%%%%%%%%%%%%%%%%%%%%%%%%%%%%%%%%%%%%%%%%%%%%%%%%%%%%%%%%%%%%%%%%%%%%%%%%%%%%%%%%%%%%%%%%

\section{Introduction}

D-branes in bosonic string theory are unstable, and their decay provides a relatively clean non-trivial example of a time-dependent background in string theory. In supersymmetric string theory, there are also
 unstable D-branes. The stable D-branes carry charges, so a pair of oppositely charged branes at subcritical
  separation also becomes unstable and will decay in a similar manner.
Investigations of properties of stable D-branes have
lead to many important insights, so one may hope that the same will be true for
unstable D-branes. The annihilation of an unstable pair of D-branes is a basic process to be understood, and also a building
block for many string-based cosmological scenarios.
However, D-brane decay is computationally a more demanding process to study. Many simple questions
remain unsolved, and have received little attention recently.\footnote{Except for the more
ambitious open string field theory program, where notable progress in tachyon condensation was
obtained in \cite{Schnabl:2005gv,Kiermaier:2007ba} and subsequent work, see \cite{Kiermaier:2008qu} for recent discussion in relating the results to boundary conformal theory.} In this paper we investigate
the amplitude for the emission of a pair of closed strings from a decaying brane, in the simple
 ``half S-brane'' rolling tachyon decay background \cite{Sen:2002nu,Larsen:2002wc}. We focus on
 D$p$-branes with $p<25$.

The single closed
string emission channel is relatively well understood, both in bosonic \cite{Lambert:2003zr,Karczmarek:2003xm,Okuyama:2003jk} and superstring \cite{Shelton:2004ij} theory, also in the
presence of background electric fields, fluxes, or winding \cite{Mukhopadhyay:2002en,Rey:2003xs,Nagami:2003yz,Gutperle:2004be}. %NJ lis??sin my??s pari boundary-state paprua: Sen ja Rey
Bulk-boundary amplitudes
(emission of a closed string from a perturbed D-brane) can also be calculated analytically
in bosonic and supersymmetric theories \cite{Balasubramanian:2004fz,Fredenhagen:2004cj,Jokela:2005ha} (see
also \cite{Hosomichi:2001xc,Jokela:2009gc}).

For multi-string amplitudes, one strategy to calculate the associated
$n$-point (bulk or boundary) amplitudes was based on previous success in bulk Liouville theory \cite{Goulian:1990qr}. Also in boundary Liouville theory, two-point and three-point functions have well-defined analytical expressions \cite{Fateev:2000ik,Teschner:2000md,Ponsot:2001ng}. In \cite{Gutperle:2003xf,Constable:2003rc} it was proposed that analytic continuation in coupling $b\rightarrow i$ and field space $\phi \rightarrow iX^0$ to timelike Liouville theory could be used to obtain results in
the rolling tachyon background. For bulk-boundary amplitudes, analytic continuation from spacelike
theory lead to a result \cite{Hosomichi:2001xc,Fredenhagen:2004cj} which is in agreement to that of \cite{Balasubramanian:2004fz} which was based on a more straightforward approach.

Closed string pair production, or associated two-point functions
were investigated in \cite{Gutperle:2003xf,Constable:2003rc}
but the obtained amplitudes did not display the expected
pole structure. Ref. \cite{Balasubramanian:2004fz} analyzed the problem
and identified some potential problems in the analytic continuation. For additional related work, see
\cite{Jokela:2008zh}.
We will comment on the analytic properties in more detail in Section \ref{sec:setup}. For the ``full S-brane''
background, an interesting prescription to compute disk $n$-point amplitudes was proposed in \cite{Gaiotto:2003rm}, and its higher order extensions were investigated in \cite{Bergman:2004pb}. It would
be interesting to investigate this method further to compare the results with those obtained on the
half S-brane background. Additional work on string scattering from decaying branes can also be found in \cite{Strominger:2002pc}.
See also the reviews on open string tachyon dynamics \cite{Sen:2004nf} and on Liouville field theory \cite{Nakayama:2004vk}.

\begin{figure}
\centering
\includegraphics[scale=0.5]{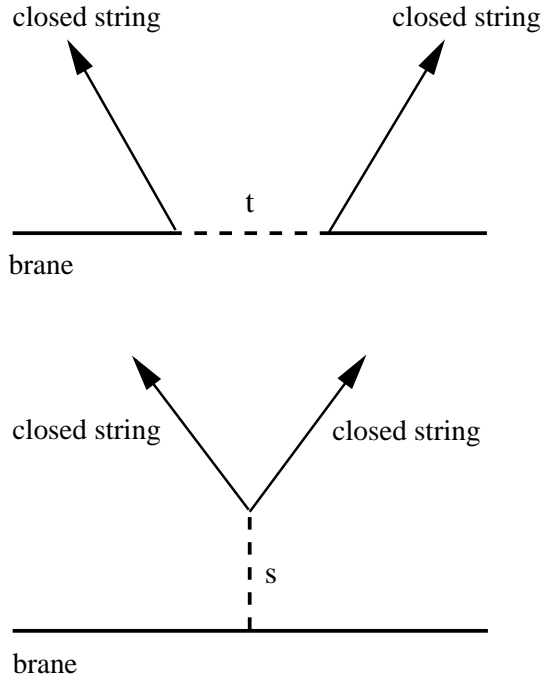}
\caption{Two channels for emission of a pair of closed strings from a decaying brane. Note the $t\leftrightarrow s$
difference compared to string scattering, see Fig. 2 of \cite{Dbrane2}.}
\label{fig:chan}
\end{figure}

In this paper we continue developing the approach initiated in \cite{Balasubramanian:2004fz}, in particular we use the relation of correlation functions in the rolling tachyon background to thermal expectation values
in a classical log gas of unit charges in two dimensions. The thermal interpretation was
further explored in \cite{Balasubramanian:2006sg,Jokela:2007dq,Jokela:2007wi,Hutasoit:2007wj,Jokela:2007yc,Schomerus:2008je}. We also use the powerful contour integral
representation which was developed in \cite{Jokela:2008zh} to study $n$-point boundary functions.
Our main result is an amplitude (obtained by working in the high-energy limit) 
for the emission of a pair of closed string tachyons from the decaying brane.
The formula differs from the previously obtained ones by its more promising pole structure. It contains the two
expected channels: an s-channel, where an on-shell closed string propagates first before splitting to the two emitted
closed strings, and
a t-channel, where a pair of on-shell open strings propagate first before closing off to the pair of closed
strings, depicted in Figure \ref{fig:chan}.

Our new calculational ingredient is the use of electrostatics to estimate the leading contribution to the correlation function. While electrostatic analogues have a long history in string theory \cite{Fairlie:1970tc},
our approach is reminiscent of that by Gross and Mende \cite{Gross:1987kza} (see also \cite{Barbon:1996ie,Bachas:1999tv}), %NJ lis??tty my??s n??it??
who used electrostatic equilibrium conditions to identify the dominant saddle point contributions to string scattering amplitudes at arbitrary
order to determine their high energy behavior. We will also be interested in the high energy limit (for the
emitted strings), but we consider only the leading disk amplitude, which is already complicated due to the
rolling tachyon background which will correspond to a thermal charged gas background.
In the high energy limit, our amplitude displays the expected exponential $\sim e^{-2\pi \omega}$ suppression, with the sum of the energies of the emitted strings in the exponent.

The paper is organized as follows. Section \ref{sec:setup} contains a brief derivation of the perturbative
series expansion of the bulk two-point amplitude. We then comment briefly on previous work and analytic properties of the series expansion,
and convert it to an integral representation appealing to \cite{Jokela:2008zh}. Section \ref{sec:statics} presents the electrostatics approach and
applies it to the emission amplitude calculation. Section \ref{sec:highenergy} introduces Mandelstam variables and presents the result for the amplitude
formula and identifies the poles. We then consider its high energy asymptotics in different kinematic limits and find that the amplitude
becomes at least exponentially suppressed, as expected. In the end there are three appendices, Appendix \ref{app:kin} collects some details on kinematics and the
amplitude calculation, for readers' convenience. Appendix \ref{app:Schur} discusses an alternative attempt to obtain an exact result for the two-point amplitude,
using a Schur polynomial method. Appendix \ref{app:genmet} contains some details of the electrostatics calculations. We also plan to present a more general discussion
of the electrostatic approximation in a forthcoming work \cite{ES}.

%%%%%%%%%%%%%%%%%%%%%%%%%%%%%%%%%%%%%%%%%%%%%%%%%%%%%%%%%%%%%%%%%%%%%%%%%%%%%%%%%%%%%%%%%%%%%%%%%%%%%%
%%%%%%%%%%%%%%%%%%%%%%%%%%%%%%%%%%%%%%%%%%%%%%%%%%%%%%%%%%%%%%%%%%%%%%%%%%%%%%%%%%%%%%%%%%%%%%%%%%%%%%
%%%%%%%%%%%%%%%%%%%%%%%%%%%%%%%%%%%%%%%%%%%%%%%%%%%%%%%%%%%%%%%%%%%%%%%%%%%%%%%%%%%%%%%%%%%%%%%%%%%%%%
%%%%%%%%%%%%%%%%%%%%%%%%%%%%%%%%%%%%%%%%%%%%%%%%%%%%%%%%%%%%%%%%%%%%%%%%%%%%%%%%%%%%%%%%%%%%%%%%%%%%%%

\section{The bulk two-point amplitude}\label{sec:setup}

We begin with setting up the
notations for different correlation functions. We work in units where $\alpha'=1$. The correlation functions are worldsheet path integrals
\be
  A_2(\{\omega_a,\vec k_a\}_{a=1}^2) = \int\prod_{a=1}^2\frac{d^2 w_a}{2\pi}\vev{\prod_{a=1}^2 V(w_a,\bar w_a)e^{-\delta S_{\rm bdry}}}_{\rm free} \ ,
\ee
where the boundary deformation representing the half S-brane \cite{Sen:2002nu,Larsen:2002wc} is
\be
 \delta S_{\rm bdry} = \la\int dt e^{X^0(t)} .
\ee
The vertex operators for the closed
string tachyons are
\be
 V(w_a,\bar w_a) = e^{i k_a\cdot X(w_a,\bar w_a)} \ ,
\ee
where $a=1,2$.
We adopt a notation $\xi_a=-i\omega_a$ and break up the spatial momentum to parallel and perpendicular directions to the unstable D$p$-brane: $\vec k_a = (\vec k_a^{\parallel},\vec k_a^\perp)$. We also denote $k^\parallel_a=(\om_a ,\vek^\parallel_a)=(i\xi_a ,\vek^\parallel_a)$. On-shell conditions for the bosonic closed string tachyons are $k_a^2=\xi_a^2+(\vec k_a)^2 = 4$.
Furthermore, one finds that the overall (spatial) parallel momentum is conserved: $\vec k^1_\parallel+\vek^2_\parallel=0$.

The worldsheet correlation functions can be evaluated by first isolating the zero modes from the oscillators, $X^\mu=x^\mu + X'^\mu$, and then
expanding the boundary deformation into a power series. This yields
\bea\label{eq:An}
  A_2(\{\omega_a,\vec k_a\}) &=& \int dx^0 d^p\vec x_\parallel e^{i\sum_a k_a^\mu x_\mu}\sum_{N=0}^\infty\frac{(-z)^N}{N!}\int\prod_{a=1}^{2}\frac{d^2 w_a}{2\pi} \nn\\
&&\times \prod_{i=1}^N\frac{dt_i}{2\pi}\vev{\prod_{i=1}^N e^{X'^0(t_i)}\prod_{a=1}^2 e^{i k_a\cdot X'(w_a,\bar w_a)}} \ ,
\eea
where we introduced $z=2\pi\lambda e^{x^0}$.

After working out the contractions, with the Neumann and Dirichlet Green's
functions (see Appendix~\ref{app:kin} and, \eg, \cite{Balasubramanian:2004fz} for more details), (\ref{eq:An}) becomes
\bea\label{eq:Anno}
  A_2(\{\omega_a,\vec k_a\}) & = & \int dx^0 d^p\vec x_\parallel e^{i\sum_a k_a^\mu x_\mu}\bar  A_2(x^0) \\
\bar A_2(x^0)
& = & \sum_{N=0}^\infty (-z)^N \int\prod_{a=1}^{2}\frac{d^2 w_a}{2\pi}|w_1-w_2|^{k_1\cdot k_2} \nonumber\\
&&\times\prod_{a, b=1}^{2}|1-w_a \bar w_b|^{\half (k^\parallel_a\cdot k^\parallel_b-\vec k^\perp_a\cdot \vec k^\perp_b)} Z_2(\{w_a,k_a\};N) \ ,
\label{eq:Abardef}
\eea
where
\bea \label{eq:int}
 Z_2(\{w_a,k_a\};N) & = & \frac{1}{N!} \int\prod_{i=1}^N\frac{dt_i}{2\pi}\prod_{1\leq i<j\leq N}|e^{it_i}-e^{it_j}|^2\prod_{a=1}^2|1-w_a e^{-it_i}|^{2\xi_a} \ .
\eea

%%%%%%%%%%%%%%%%%%%%%%%%%%%%%%%%%%%%%%%%%%%%%%%%%%%%%%%%%%%%%%%%%%%%%%%%%%%%%%%%%%%%%%%%%%%%%%%%%%%%%%
%%%%%%%%%%%%%%%%%%%%%%%%%%%%%%%%%%%%%%%%%%%%%%%%%%%%%%%%%%%%%%%%%%%%%%%%%%%%%%%%%%%%%%%%%%%%%%%%%%%%%%
%%%%%%%%%%%%%%%%%%%%%%%%%%%%%%%%%%%%%%%%%%%%%%%%%%%%%%%%%%%%%%%%%%%%%%%%%%%%%%%%%%%%%%%%%%%%%%%%%%%%%%
%%%%%%%%%%%%%%%%%%%%%%%%%%%%%%%%%%%%%%%%%%%%%%%%%%%%%%%%%%%%%%%%%%%%%%%%%%%%%%%%%%%%%%%%%%%%%%%%%%%%%%

\subsection{Some general comments} \label{sec:general}

We fix the residual global conformal symmetry on the disk by placing the other bulk vertex operator to the origin ($w_2=0$) and the other one to the real axis at $1>w_1=r>0$. Let the  (imaginary) energy of the closed string vertex %charge
at $w_1=r$ be $\xi_1\equiv \xi$ and the energy of the string at $w_2=0$ be $\xi_2$. The first challenge
 in calculating the amplitude is the integrals (\ref{eq:int}), which now become
\be \label{eq:Z2def}
 Z_{2}(r,\xi;N) =  \inv{N!}\int\left[\prod_{i=1}^{N}\frac{dt_i}{2\pi}|r-e^{it_i}|^{2\xi}\right]
 \left[\prod_{1\leq i<j\leq N}|e^{it_i}-e^{it_j}|^{2}\right] \ .
\ee
We can make the following general remarks:
\begin{itemize}
 \item For $0 \le r<1$ the integral $Z_{2}(r,\xi;N)$ is an analytic function of $\xi$.
 \item For negative integer $\xi$ and for $N \ge |\xi|$ the integral can be evaluated by using identities of the Schur polynomials \cite{Constable:2003rc} giving $Z_{2}(r,\xi;N) = (1-r^2)^{-\xi^2}$. For positive integer $\xi$ the integral is a polynomial in $r$, and may be written down explicitly by using Schur polynomials (see Appendix~\ref{app:Schur}).
 \item In the limit $N \to \infty$ authors of \cite{Balasubramanian:2004fz} also find $Z_{2}(r,\xi;N) \to (1-r^2)^{-\xi^2}$.
 \item For $r \to 1$, and if $\re \xi>-1/2$,\footnote{Authors of \cite{Balasubramanian:2004fz} use functional relations which hold only for $\re\xi>-1/2$ in their discussion of the $r \to 1$ limit.} $Z_{2}(r,\xi;N)$ approaches smoothly the boundary-one-point and bulk-boundary functions \cite{Balasubramanian:2004fz} (see also \cite{Fredenhagen:2004cj}).
\end{itemize}
In the final amplitude the interesting region is $\xi=-i\omega$, where $\omega$ is the energy of the closed string at $w_1=r$. In the case of (integrated) boundary amplitudes, it is possible to analytically continue from positive $\re\xi$ to imaginary values \cite{Jokela:2008zh}.
The bulk integrals approach smoothly the boundary ones for positive $\re\xi$ in the limit $r \to 1$, so it is natural to assume that the $\re\xi>0$ region may be a good starting point to study
the complete bulk amplitudes.

There are closely related problems where the scattering amplitude, the sum of $N$ over an
infinite series of terms, is essentially equal to a generic $N$th term after analytically
continuing $N\rightarrow -i\sum_a\omega_a$. In spacelike Liouville theory, for bulk amplitudes
a related observation was made in \cite{Zamolodchikov:1995aa}. In \cite{Balasubramanian:2004fz} it was observed for the bulk-boundary
amplitude in the decaying D-brane background. Then, ref. \cite{Jokela:2008zh} constructed a detailed contour integration method to show how boundary amplitudes reduce to the analytical continuation
of a term. Following \cite{Jokela:2008zh,Balasubramanian:2004fz}
we expect %\todo{can we say this more strongly?}
that the bulk amplitude is given by a certain analytic continuation to negative $N$:
\beq \label{eq:A2def}
 A_{2} = \frac{\pi(2\pi\la)^{-\xi-\xi_2}}{\sin\pi(\xi+\xi_2)} \int_0^1 dr r r^{k_1\cdot k_2} (1-r^2)^{\inv{2}\left((k_1^{||})^2-(\vek_1^{\perp})^2 \right)} Z_{2}(r,\xi;N=-\xi-\xi_2) \ .
\eeq

In the integrand of (\ref{eq:A2def}), at this point we only know the term
$Z_{2}$ through its definition in \eq{Z2def}, which is quite complicated. However, it allows
us to make some observations about the pole structure of the amplitude. There
singularities at each endpoint $r=0,1$ in the $r$ integral give rise to pole structures,
as in scattering from a stable D-brane \cite{Dbrane1},
see \cite{Dbrane2} for a review.\footnote{On the other hand, scattering from a stable D-brane corresponds to the $N=0$ term in the series expansion (\ref{eq:An}).} From \eq{Z2def}
we can see that $Z_2$ is real analytic at $r=0$ and even in $r$. It is also analytic at $r=1$ for $\re \xi>-1/2$. Consequently, the singularities at the endpoints originate from the factor $ r r^{k_1\cdot k_2} (1-r^2)^{\inv{2}\left((k_1^{||})^2-(\vek_1^{\perp})^2 \right)} $, which is the same as in scattering from the stable D-brane. Let us consider the poles from $r=0$ in more detail.
We may expand $Z_{2}$ as a Taylor series at $r=0$,
\bea
 Z_2(r,\xi;N\!=\!-\xi\!-\!\xi_2) &=& Z_2(0,\xi;N\!=\!-\xi\!-\!\xi_2) + \inv{2} r^2 \frac{\partial^2}{\partial r^2}\!\!\left.Z_2(r,\xi;N\!=\!-\xi\!-\!\xi_2)\right|_{r=0} \! + \cdots \nn \\
              &=& 1 + a_2(\xi,\xi_2) r^2  + \cdots \ ,
\eea
where $a_2(\xi,\xi_2)$ is the proper analytic continuation of the series coefficient\footnote{See Appendix~\ref{app:Schur} for a calculation of the series coefficients for $N=0,1,2,\ldots$. In particular, for integer $N$ one can extract from eqn. \eq{Z2intN} that $a_2 = \xi^2\delta_{N>0}$.} to $N=-\xi-\xi_2$.
Integrating over $r$ term by term in \eq{A2def} then yields
\bea
 A_2  &=&  \frac{\pi(2\pi\la)^{-\xi-\xi_2}}{\sin\pi(\xi+\xi_2)} \int_0^1\!\! dr r r^{k_1\cdot k_2}\!\left[1 + \left(a_2(\xi,\xi_2)-\inv{2}(k_1^{||})^2 + \inv{2}(\vek_1^{\perp})^2 \right) r^2  + \cdots\right] \nn \\
      &=& \frac{\pi(2\pi\la)^{-\xi-\xi_2}}{\sin\pi(\xi+\xi_2)} \left[ \inv{k_1\cdot k_2+2} + \frac{a_2(\xi,\xi_2)-\inv{2}(k_1^{||})^2 + \inv{2}(\vek_1^{\perp})^2 }{k_1\cdot k_2+4} + \cdots \right] \ .
\eea
The poles are located at $k_1\cdot k_2 = -2,-4,\ldots$, as in scattering from ordinary D-branes \cite{Dbrane1,Dbrane2}. The above calculation highlights the structure of the integrand at $r=0$.

Similarly one can check that the singularity at $r=1$ gives rise to poles at $ 2 (k_1^{\pp})^2 = 2,0,-2,\ldots $ where we used the mass-shell condition $k_1^2 = 4$.

The remaining paper focuses on developing an electrostatic approximation for $Z_{2}$. It is analytic at $r=0,1$, so
the poles indeed arise as described above, and we can verify the expressions explicitly.
We shall discuss in detail only the case $\xi=\xi_2$ where the production amplitude is symmetric.\footnote{The reason for this restriction is a puzzling observation for which we do not have a good explanation at the moment.
When we use conformal symmetry to place the other vertex operator to origin and the other one to $w=r$, the
amplitude (\ref{eq:Anno})-(\ref{eq:int}) becomes asymmetric with respect to the exchange $1 \leftrightarrow 2$ of the strings, if they have different energies, $\xi \ne \xi_2$. As a result, the poles seem not to respect the exchange symmetry.
So, the use of conformal symmetry to fix the vertex positions in the rolling tachyon background involves a subtlety, unless $\xi=\xi_2$, in which case the strings are exchange symmetric ``in the time direction''.
}

%%%%%%%%%%%%%%%%%%%%%%%%%%%%%%%%%%%%%%%%%%%%%%%%%%%%%%%%%%%%%%%%%%%%%%%%%%%%%%%%%%%%%%%%%%%%%%%%%%%%%%
%%%%%%%%%%%%%%%%%%%%%%%%%%%%%%%%%%%%%%%%%%%%%%%%%%%%%%%%%%%%%%%%%%%%%%%%%%%%%%%%%%%%%%%%%%%%%%%%%%%%%%
%%%%%%%%%%%%%%%%%%%%%%%%%%%%%%%%%%%%%%%%%%%%%%%%%%%%%%%%%%%%%%%%%%%%%%%%%%%%%%%%%%%%%%%%%%%%%%%%%%%%%%
%%%%%%%%%%%%%%%%%%%%%%%%%%%%%%%%%%%%%%%%%%%%%%%%%%%%%%%%%%%%%%%%%%%%%%%%%%%%%%%%%%%%%%%%%%%%%%%%%%%%%%

\section{Electrostatic calculation} \label{sec:statics}

Our starting point is to view \cite{Balasubramanian:2004fz,Jokela:2005ha,Balasubramanian:2006sg,Jokela:2008zh}
$Z_2$ as the partition function of a Coulomb gas \cite{Dyson:1962es},
with two additional external charges $\xi$, $\xi_2$ at fixed positions.  We write it
in the form
\be \label{eq:Zredef}
 Z_2(\{\xi_a\};N)  = \frac{1}{N!} \int\prod_{i=1}^N\frac{dt_i}{2\pi} e^{-2 H}\ ,
\ee
where the Hamiltonian $H$ reads
\be \label{eq:Ham}
 H = - \sum_{1\le i<j\le N} \log |e^{it_i}-e^{it_j}| - \sum_{i=1}^N\sum_{a=1}^2 \xi_a \log |e^{it_i}-w_a|  \ ,
\ee
and we fixed the inverse temperature to $\beta=2$. As seen from the Hamiltonian, we have an ensemble of $N$ unit charges on the circle at $e^{it_1},\ldots,e^{it_N}$, which interact via the Coulomb potential.\footnote{Other unstable D-brane backgrounds correspond to different Coulomb gases \cite{Jokela:2007dq,Jokela:2007wi,Hutasoit:2007wj}.}
They are subject to an external electric field created by two fixed charges.

A useful strategy is to study the system using classical electrostatics, first in the limit of large number $N$ of unit charges. In this limit the unit charges can be
treated as a continuous charge distribution. Its density is modified by the external charges $\xi_a$ and we are interested in finding the electrostatic
equilibrium configuration. Notice that the present system is analogous to the circular unitary ensemble \cite{Larsen:2002wc} of random matrix theory, where
the connection to the Coulomb gas is a standard tool. Our method indeed resembles the electrostatic derivation of Wigner's semi-circle
law \cite{wigner} (see also \cite{Forrester}). It is also reminiscent of the work of Gross and Mende \cite{Gross:1987kza}, who used electrostatic equilibrium
conditions to identify the dominant saddle point contributions to string scattering amplitudes at arbitrary
order to determine their high energy behavior. We will also consider the high energy limit (for the
emitted strings), but we restrict to the leading disk amplitude, which is already complicated due to the charge density
background. A more detailed description of our method and other applications will be discussed elsewhere \cite{ES}, here we focus on
aspects that are relevant for the bulk two-point amplitude.

We use a saddle point approach and
write
\be \label{eq:esappr}
 \log Z_2(\{\xi_a\};N) \simeq - 2 H_0    \simeq - 2 \mE_{\rm tot} \ ,
\ee
where $H_0$ is the minimum value of the Hamiltonian \eq{Ham} and $\mE_{\rm tot}$ is the (total) electrostatic energy of the corresponding equilibrium configuration with continuous charge densities.
The string production amplitude \eq{A2def} requires $N=-\xi-\xi_2$, so that we need
to study the electrostatic problem in the regime $N \sim \xi \sim \xi_2$. It turns out that
nontrivial electrostatic configurations are indeed obtained when the external charges are of the same order as the sum of the unit charges.
Slightly more precisely, the electrostatic configurations of interest to us are in the limit $N \to \infty$ with the ratios $N/\xi$ and $N/\xi_2$ fixed.\footnote{Recall that the behavior of  $Z_2$ in the limit $N \to \infty$ with $\xi_a$ and $w_a$ fixed can be calculated by using the asymptotics of Toeplitz determinants \cite{Balasubramanian:2004fz,Fisher,Widom,Basor}.}

Let us then discuss how the total electrostatic energy $\mE_{\rm tot}$ is calculated. The first
ingredient is the (complex) potential $V(w)$, the physical potential is the real part
\be
 U (w) = \re  V (w) \ .
\ee
The potential is sourced by the continuous charge density $\rho (\phi)$
(by which we approximate the $N$ unit charges in the $N\rightarrow \infty$ limit, with the
normalization $N=\oint d\phi \rho (\phi)$) and the external
charges $\xi_a$,
\be \label{eq:Vdef}
 V(w) = - \int d\phi\rho(\phi) \log\left(w-e^{i\phi}\right) - \sum_a \xi_a \log(w-w_a) + {\rm const.} \ .
\ee
We choose the zero level of the potential by setting the last constant term to zero. We then need to
find $V(w)$ and $\rho (\phi)$ in the equilibrium configuration. We are interested in the configurations where $\rho(\phi)$ is strictly positive within a domain $C$ of the unit circle and vanishes elsewhere.\footnote{In this paper it is sufficient to assume that the domain $C$ of non-zero $\rho (\phi)$ is connected, that is, an arc.} Then solving the equilibrium configuration reverts to a standard potential problem of classical electrostatics including  the external charges $\xi_a$ and a conductor which fills the domain $C$.
In particular, in the equilibrium configuration the physical potential
$U(e^{i\phi})$ must take a constant
value,
\be
 U (e^{i\phi}) = U_0 \ ,
\ee
in $C$: If the potential would vary within $C$, the charges on the conductor would be subject to forces and move until the potential
adjusts to a constant value.

The electrostatic energy is defined as the energy of the system in the equilibrium configuration, and may be expressed in terms of $\rho_0(\phi)$ as
\bea \label{eq:Etotdef}
 \mE_{\rm tot} &=& -\inv{2}\int d\phi_1 d\phi_2 \rho_0(\phi_1) \rho_0(\phi_2) \log\left|e^{i\phi_1}-e^{i\phi_2}\right| \nn\\
  && - \sum_a \xi_a \int d\phi\rho_0(\phi) \log\left|e^{i\phi}-w_a\right| \nn \\
  &\equiv& \mE_{c} + \mE_{c-\xi} \ .
\eea
We again set a possible constant contribution to zero. Then the energy matches with the continuum limit of the Hamiltonian \eq{Ham} and thus approximates the saddle point value of $-\half \log Z_2$. By using the fact that the physical potential $U(e^{i\phi})$ is constant within the domain $C$ of nonzero $\rho_0$, we can express the result as
\be \label{eq:Etotres}
 \mE_{\rm tot}  = \frac{N}{2} U_0 + \inv{2}\mE_{c-\xi} \ ,
\ee
where the first term includes one half of the ``interaction energy'' $\mE_{c-\xi}$.
Hence, to calculate $\mE_{\rm tot}$ and the leading behavior of $Z_2$ for large $N \sim \xi \sim \xi_2$, it is sufficient to find $U_0$ and $\mE_{c-\xi}$. We present a general method for this in Appendix~\ref{app:genmet}.

In the next two subsections we
work out the electrostatics approximation of the (two-point) bulk amplitude explicitly.
We can omit the charge $\xi_2$ at the origin since it does not affect the calculation. The treatment of the other
(real) charge $\xi$ at the real line with $0<w_1=r<1$  is divided to two regimes,
depending on the values of $r$ and $\xi/N$.
For small values of $r$ the charge $\xi$ affects the charge distribution on the unit circle less, and it remains strictly positive for all $\phi=\arg w$.
For $r \to 1$ (assuming $\xi>0$) the charge $\xi$ creates a gap in the charge distribution on the circle in the interval $-\phi_c < \phi <\phi_c $.\footnote{The distribution remains strictly positive
if $\xi$ is negative and $N>|\xi|$. We shall consider the gap creation only for positive $\xi$ where the $r \to 1$ limit of $Z_2$ matches with the one-point boundary function \cite{Balasubramanian:2004fz}.} We discuss first the small $r$ case.

%%%%%%%%%%%%%%%%%%%%%%%%%%%%%%%%%%%%%%%%%%%%%%%%%%%%%%%%%%%%%%%%%%%%%%%%%%%%%%%%%%%%%%%%%%%%%%%%%%%%%%
%%%%%%%%%%%%%%%%%%%%%%%%%%%%%%%%%%%%%%%%%%%%%%%%%%%%%%%%%%%%%%%%%%%%%%%%%%%%%%%%%%%%%%%%%%%%%%%%%%%%%%
%%%%%%%%%%%%%%%%%%%%%%%%%%%%%%%%%%%%%%%%%%%%%%%%%%%%%%%%%%%%%%%%%%%%%%%%%%%%%%%%%%%%%%%%%%%%%%%%%%%%%%
%%%%%%%%%%%%%%%%%%%%%%%%%%%%%%%%%%%%%%%%%%%%%%%%%%%%%%%%%%%%%%%%%%%%%%%%%%%%%%%%%%%%%%%%%%%%%%%%%%%%%%

\subsection{Small $r$}

For small $r$ one needs to solve the potential problem on the unit disk (without gaps), which is easily done by using a mirror charge.
One can check that $U(e^{i\phi})=U_0=0$ in this case when the constant term in \eq{Vdef} is set to zero.
At electrostatic equilibrium, the potential becomes
\bea
 V(w)   &=& -\xi \log\frac{w-r}{1 - w r} \ .
\eea
The first part of \eq{Vdef}, the contribution from the continuous distribution on the circle, becomes $V_c(w) =  +\xi \log(1-wr)$ at equilibrium, which in turn equals the potential of the mirror charge at $w=1/r$. As mentioned above, we dropped the trivial contribution $-\xi_2\log w$ due to the charge at the origin which neither affects the charge densities nor the total energy. The equilibrium charge distribution on the unit circle is
\be
  \rho_0(\phi) = \inv{2\pi}\left(N+\xi\right) - \inv{2\pi}\left|V'(w)\right|_{w=e^{i \phi}} = \inv{2\pi}\left(N+\xi-\xi\frac{1-r^2}{1+r^2-2r \cos\phi}\right) \ ,
\ee
where the first (constant) term is fixed by $\oint d\phi\rho(\phi) = N$ and the second ($\phi$ dependent) term is the charge density induced by the charge $\xi$ at $w_1=r$.
The absence of the gap requires that $\rho$  is positive, \ie, that
\be
\label{eq:gap}
 r < \frac{N}{N+2\xi} \equiv r_c \ ,
\ee
which defines what we mean by ``small $r$''.
The total energy at equilibrium becomes
\bea \label{eq:Etotsr}
 \mE_{\rm tot} &=& \inv{2}\mE_{c-\xi} = \frac{\xi}{2} V_c(w=r) = \frac{\xi^2}{2} \log\left(1-r^2\right) \ .
\eea
Notice that in the large $N$ limit with fixed $\xi$ we have $r_c\to 1$, and the result \eq{Etotsr} is valid for all $r$. Hence in this limit we may use the result directly in \eq{esappr},
\be
 Z_2(\xi;N) \propto e^{-2 \mE_{\rm tot}} = \left(1-r^2\right)^{-\xi^2} \ ; \qquad N \to \infty \ ,
\ee
which indeed reproduces exactly the large $N$ result of eqn. (5.13) in \cite{Balasubramanian:2004fz}. However, as we will
see next, at large $r$ we obtain in general a different result.

%%%%%%%%%%%%%%%%%%%%%%%%%%%%%%%%%%%%%%%%%%%%%%%%%%%%%%%%%%%%%%%%%%%%%%%%%%%%%%%%%%%%%%%%%%%%%%%%%%%%%%
%%%%%%%%%%%%%%%%%%%%%%%%%%%%%%%%%%%%%%%%%%%%%%%%%%%%%%%%%%%%%%%%%%%%%%%%%%%%%%%%%%%%%%%%%%%%%%%%%%%%%%
%%%%%%%%%%%%%%%%%%%%%%%%%%%%%%%%%%%%%%%%%%%%%%%%%%%%%%%%%%%%%%%%%%%%%%%%%%%%%%%%%%%%%%%%%%%%%%%%%%%%%%
%%%%%%%%%%%%%%%%%%%%%%%%%%%%%%%%%%%%%%%%%%%%%%%%%%%%%%%%%%%%%%%%%%%%%%%%%%%%%%%%%%%%%%%%%%%%%%%%%%%%%%

\subsection{Large $r$}

As we saw in (\ref{eq:gap}), for $r>N/(N+2\xi)$ there is a gap on the unit circle charge distribution.
The electrostatic problem becomes more difficult.  We present a general method to find the equilibrium
charge density in the presence of gaps in Appendix~\ref{app:genmet}. Here we focus on the results in the case of one gap $\hat n=1$ and one external charge $n=1$ which is relevant at large $r$.

Appendix~\ref{app:genmet} performs the calculation in a simplified geometry, where the unit disk (in $w$ coordinates)
is mapped onto the upper half plane ($z$ coordinates) by
\be \label{eq:confmap2}
 z = q(w) = i \beta \frac{1-w}{1+w} = i \cot\left(\phi_c/2\right)\, \frac{1-w}{1+w} \ ,
\ee
where $\phi_c$
will parameterize the location of the gap in the charge density.
In the electrostatic problem (on the $w$ plane),
there is an induced image charge $-N-\xi$ at infinity, so that the net charge is zero.
Following the point at infinity in the mapping $q$ in \eq{confmap2}, the image appears as an additional charge at $z=q(\infty)$ on the $z$ plane.
In the notation of Appendix~\ref{app:genmet} the actual charge is
$\xi_n=\xi_1 \equiv \xi$ and the image charge is $\xi_{n+1}=\xi_2$:
\beqa
 \xi_1 = \xi \quad &\mathrm{at}& \quad z=i\delta(r) \beta  \quad (w=r) \nn\\
 \xi_2 = - N- \xi \quad &\mathrm{at}& \quad z=-i\beta \quad (w=\infty) \ ,
\eeqa
where $\beta= \cot(\phi_c/2)$ and
\be
 \delta(r) = \frac{1-r}{1+r} \ .
\ee
The asymptotic behavior of the electric field leads to a constraint
(eqn. \eq{constraints2} in Appendix~\ref{app:genmet}) which gives a relation
\be
 \left(1+\frac{N}{\xi}\right)^2 = \frac{1+\delta^2(r) \beta^2}{1+\beta^2}
\ee
or equivalently
\be
 \cos^2\frac{\phi_c}{2} = \frac{1-\chi^2}{1-\delta(r)^2} = \frac{1-\left(\frac{\xi}{N+\xi}\right)^2}{1-\left(\frac{1-r}{1+r}\right)^2} \ ,
\ee
where $\chi= \xi/(N+\xi)$. Notice that $\phi_c$ vanishes at $r=r_c=N/(N+2\xi)$ which signals the disappearance of the gap.
By using eqs.~\eq{VVcdef} and \eq{Udef} we find for the constant value of the potential
\bea \label{eq:Vcbulk}
 U_0
 &=& -\frac{N}{2}\log \frac{1-\chi}{1-\delta(r)} - \frac{N+2 \xi}{2}\log \frac{1+\chi}{1+\delta(r)} \ ,
\eea
while the interaction energy of \eq{Ecxidef} is given by
\bea \label{eq:Exibulk}
 \mE_{c-\xi}
&=& \xi(N+\xi)\left[\log\frac{1+\delta(r)}{1+\chi}+\chi\log\frac{4\chi}{(1+\chi)(1+\delta(r))}\right] \ .
\eea
The total energy thus reads
\bea \label{eq:Etotlr}
 \mE_{\rm tot} &=& \frac{N}{2}U+\frac{1}{2}\mE_{c-\xi} \nn\\
&=&  - \frac{(N+2 \xi)^2}{4}\log \frac{1+\chi}{1+\delta(r)}-\frac{N^2}{4}\log \frac{1-\chi}{1-\delta(r)}+\frac{\xi^2}{2}\log\frac{4\chi}{(1+\delta(r))^2} \ .
\eea

%%%%%%%%%%%%%%%%%%%%%%%%%%%%%%%%%%%%%%%%%%%%%%%%%%%%%%%%%%%%%%%%%%%%%%%%%%%%%%%%%%%%%%%%%%%%%%%%%%%%%%
%%%%%%%%%%%%%%%%%%%%%%%%%%%%%%%%%%%%%%%%%%%%%%%%%%%%%%%%%%%%%%%%%%%%%%%%%%%%%%%%%%%%%%%%%%%%%%%%%%%%%%
%%%%%%%%%%%%%%%%%%%%%%%%%%%%%%%%%%%%%%%%%%%%%%%%%%%%%%%%%%%%%%%%%%%%%%%%%%%%%%%%%%%%%%%%%%%%%%%%%%%%%%
%%%%%%%%%%%%%%%%%%%%%%%%%%%%%%%%%%%%%%%%%%%%%%%%%%%%%%%%%%%%%%%%%%%%%%%%%%%%%%%%%%%%%%%%%%%%%%%%%%%%%%

\section{High energy behavior}\label{sec:highenergy}

Let us then apply the electrostatic results to estimate the string production amplitude of \eq{A2def}.
In the bulk case there is a subtlety with analytic continuation of \cite{Balasubramanian:2004fz,Jokela:2008zh} $N \to -\xi-\xi_2$.
When $r$ is kept fixed, the integral can be approximated by the ``large $r$'' result \eq{Etotlr} for $N \lesssim N_c$, where $N_c = 2\xi r/(1-r)$, whereas ``small $r$'' approximation \eq{Etotsr} works for $N\gtrsim N_c$. Thus there is a discontinuity at $N=N_c$, which leads to apparent problems with the analytic continuation. The sum that we should calculate reads
\be
 \bar A_{2}(z) \sim \sum_{N=0}^{\infty} (-z)^N Z_2(r,\xi;N) \simeq \sum_{N=0}^{\infty} (-z)^N e^{-2 \mE_{\rm tot}} \ ,
\ee
where $z=2\pi \la e^{x^0}$ and $\mE_{\rm tot}$ is given by either \eq{Etotlr} if $N\lesssim N_c$ or \eq{Etotsr} if $N\gtrsim N_c$.
The sum is well convergent and insensitive to the actual value of $N_c$ for $|z|<1$. For continuation to $|z|>1$ we would like to apply the contour integral method of \cite{Jokela:2008zh}. However, the discontinuity at $N=N_c$ seems to be problematic, if $r$ is held fixed. The sum $\bar A_2(z)$ is sensitive to $N_c$ and  consequently diverges fast for $z \to \infty$ with an arbitrary proportionality constant, whence it is not possible to integrate the amplitude over $x^0$. Working directly at imaginary $\xi=-i\omega$ might improve the convergence, but the analytic continuation is still tricky.

Remarkably, the problems disappear if $r$ is integrated over first. Hence we define the integrated partition function
\bea
  \int_0^1 dr k(r) Z_2(r,\xi;N)
&\simeq& \int_0^{r_c} dr k(r) \left(1-r^2\right)^{-\xi^2} + \int_{r_c}^1 dr k(r) \left(\frac{1+\chi}{1+\delta(r)}\right)^{(N+2\xi)^2/2} \nonumber\\
 && \times \left(\frac{1-\chi}{1-\delta(r)}\right)^{N^2/2}  \left(\frac{4 \chi}{(1+\delta(r))^2}\right)^{-\xi^2} \nonumber \\
 &\equiv&  \bar Z(\xi;N) \ ,
\eea
where $r_c=N/(N+2\xi)$ and $k(r)$ can be any arbitrary analytic function, but we will choose it appropriately below in \eq{kdef}. Now $\bar Z$ is an analytic function of $N$, and can be continued analytically to all complex $N$ plane. It has a few branch cuts: branch points are found whenever the endpoint of the integration, $r_c$, hits singularities of the integrand. Branches are given by the various winding possibilities of the integration path. The singularities are found at $r=0$, $\pm 1$, and $\infty$, giving for the branch points $N=0$, $-\xi$, $-2\xi$, and $\infty$.

It is straightforward to check that  $\bar Z$ grows only as a power law as $N \to \infty$ in any direction on the complex plane. Hence, following the analysis of \cite{Jokela:2008zh}, we can apply the contour integration method, which gives our conjecture for the high energy behavior of the bulk two-point function
\be
 A_{2}  \simeq  \int dx^0\, e^{(\xi+\xi_2)x^0} \sum_{N=0}^{\infty} (-z)^N \bar Z(N) = \frac{\pi(2\pi\la)^{-\xi-\xi_2}}{\sin\pi(\xi+\xi_2)} \bar Z (\xi;N=-\xi-\xi_2) \ ,
\ee
where we fixed (see \eq{A2def})
\be \label{eq:kdef}
 k(r) = r r^{k_1\cdot k_2}(1-r^2)^{\inv{2}\left((k_1^{||})^2-(\vek_1^{\perp})^2 \right)}
\ee
and omitted the trivial factor $\delta\left(\vek_1^{||}+\vek_2^{||}\right)$.
Explicitly,
\bea \label{eq:bulk2pt}
 A_{2} & \simeq &\frac{\pi(2\pi\la)^{-\xi-\xi_2}}{\sin\pi(\xi+\xi_2)}\Bigg[\int_0^{\hat r_c} dr  r^{k_1\cdot k_2+1}  \left(1-r^2\right)^{\inv{2}\left(-\xi^2+(\vek_1^{||})^2-(\vek_1^{\perp})^2 \right)}  \\
&&+ \int_{\hat r_c}^1 dr  r^{k_1\cdot k_2+1}  \left(1-r^2\right)^{\inv{2}\left((k_1^{||})^2-(\vek_1^{\perp})^2 \right)}\nn\\
&&\times  \left(\frac{(1+\delta(r))^2}{4\hat \chi}\right)^{\xi^2} \left(\frac{1+\hat \chi}{1+\delta(r)}\right)^{(\xi-\xi_2)^2/2}  \left(\frac{1-\hat \chi}{1-\delta(r)}\right)^{(\xi+\xi_2)^2/2} \Bigg] \ ,\nonumber
\eea
where the hats indicate that analytic continuation was imposed:
\be
 \hat r_c = \frac{\xi_2+\xi}{\xi_2-\xi}\ , \qquad \hat  \chi = -\frac{\xi}{\xi_2}\ , \quad \mathrm{and} \quad \delta(r) = \frac{1-r}{1+r} \ .
\ee
Here we may use the momentum conservation in (spatial) directions parallel to the D$p$-brane and the on-shell conditions, which lead to
\bea \label{eq:kincons}
 \vek_1^{||} &=& -\vek_2^{||} \equiv \vek^{||} \nonumber \\
 (\vek_1^\perp)^2+\xi^2 &=& (\vek_2^\perp)^2+\xi_2^2 = 4 - (\vek^{||})^2 \ .
\eea

Notice that the result seems to be asymmetric under the reordering $1\leftrightarrow 2$ of the vertices unless $\xi =\xi_2$ as suggested by the analysis of the pole positions in Section~\ref{sec:general}. In general $\xi \ne \xi_2$ unless the brane is space-filling ($p=25$). We will now focus on the symmetric case $\xi=\xi_2$, which leads to a drastic simplification of the result \eq{bulk2pt}. By \eq{kincons} this then also sets $|\vek_1^\perp|=|\vek_2^\perp|$. Let us define the kinematic variables
\bea
 s &=& k_1\cdot k_2 \nn\\
 t &=& 2 (k_1^{||})^2 = 2 (k_2^{||})^2 = 8 - (\vek_1^{\perp})^2-(\vek_2^{\perp})^2 \nn\\
 u &=& \frac{(\vek_1^{\perp}-\vek_2^{\perp})^2}{2} \ ,
\eea
which are motivated by the standard conventions (see Appendix~\ref{app:kin}) for scattering from a stable D-brane \cite{Dbrane1,Dbrane2} but should not be regarded as true Mandelstam variables.\footnote{These variables are not directly linked to $2 \to 2$ scattering which is the case for the stable D-brane (Appendix~\ref{app:kin}). For example, even after fixing $\xi=\xi_2$ there are three independent kinematic variables in the present case. Hence $u$ cannot be expressed in terms of $s$ and $t$ in contrast to usual $2 \to 2$ scattering.} We have crossed $s \leftrightarrow t$ to obtain a setting which corresponds to string production by the brane rather than scattering off the brane.

For equal energies \eq{bulk2pt} is apparently singular, because $N$ is continued analytically to the value $-2\xi$ where the integral has a branch point. Hence we need to check the details of the analytic continuation carefully. We choose a natural branch where the continuation is done from positive values of $N$ via the path
\beq
 N = 2\xi e^{i\varphi} \ \mathrm{;} \qquad \varphi=0\ldots \pi \ .
\eeq
Then, in particular, $\hat r_c \to \infty$ along the positive imaginary axis, or more precisely the path $\re\, \hat r_c=1/2$.
The amplitude becomes
\bea \label{eq:bulk2pt2}
 A_{2} & \simeq &\frac{\pi(2\pi\la)^{-2\xi}}{2\sin2\pi\xi}\Bigg\{e^{i\pi\xi^2}\int_0^1 dx  x^{\frac{s}{2}-\xi^2}(1-x)^{\frac{t}{2}-2} \nn\\
&& - e^{i\frac{\pi s}{2}} \int_0^\infty dy\left[y^{\frac{s}{2}-\xi^2}(1+y)^{\frac{t}{2}-2}-y^{\frac{s}{2}}(1+y)^{\frac{t}{2}-\xi^2-2}\right] \Bigg\} \ ,
\eea
where we substituted $x=r^2$ and $y=-r^2$.
After using identities of the gamma functions, the result simplifies to
\bea \label{eq:bulk2ptfin}
 A_{2} & \simeq &-\frac{ i \pi e^{i\frac{\pi s}{2}}(2\pi\la)^{2 i\omega}}{2\sinh2\pi\omega}\left[e^{i\frac{\pi u}{2}}B\Big(\frac{t}{2}-1,\frac{u}{2}-1\Big)+B\Big(\frac{s}{2}+1,\frac{u}{2}-1\Big)\right] \nn\\
& = &-\frac{ i \pi e^{i\frac{\pi s}{2}}(2\pi\la)^{2 i\omega}}{2\sinh2\pi\omega}\left[e^{i\frac{\pi u}{2}}\frac{\Gamma\left(\frac{t}{2}-1\right)}{\Gamma\left(\frac{t+u}{2}-2\right)}+\frac{\Gamma\left(\frac{s}{2}+1\right)}{\Gamma\left(\frac{s+u}{2}\right)}\right]\Gamma\left(\frac{u}{2}-1\right) \ ,
\eea
where $\omega=\omega_1=\omega_2=i \xi$.

Notice that \eq{bulk2ptfin} has the expected $s$- and $t$-channel poles at $s = -2,-4,-6,\ldots $ and at  $t = 2,0,-2,\ldots$ which arise from the endpoints at $r=0$ and $r=1$ of the integral in \eq{bulk2pt}, respectively.
It is possible to check that the residues of the first few poles are in accord with the general expectations of Section~\ref{sec:general}.
The last gamma function produces an additional pole at $u=0$, while at $u=2$ the expression in square brackets vanishes which cancels the next pole. The following poles at negative $u$ are unphysical since $u\ge 0$ by definition. The $u=0$ singularity corresponds to production of strings which have parallel momenta in the directions perpendicular to the D$p$-brane, $\vek_1^\perp=\vek_2^\perp$, $p<25$. The amplitude is also singular at $\omega = 0$.

%%%%%%%%%%%%%%%%%%%%%%%%%%%%%%%%%%%%%%%%%%%%%%%%%%%%%%%%%%%%%%%%%%%%%%%%%%%%%%%%%%%%%%%%%%%%%%%%%%%%%%
%%%%%%%%%%%%%%%%%%%%%%%%%%%%%%%%%%%%%%%%%%%%%%%%%%%%%%%%%%%%%%%%%%%%%%%%%%%%%%%%%%%%%%%%%%%%%%%%%%%%%%
%%%%%%%%%%%%%%%%%%%%%%%%%%%%%%%%%%%%%%%%%%%%%%%%%%%%%%%%%%%%%%%%%%%%%%%%%%%%%%%%%%%%%%%%%%%%%%%%%%%%%%
%%%%%%%%%%%%%%%%%%%%%%%%%%%%%%%%%%%%%%%%%%%%%%%%%%%%%%%%%%%%%%%%%%%%%%%%%%%%%%%%%%%%%%%%%%%%%%%%%%%%%%

\subsection{Kinematic limits}

\begin{figure}
\centering
\includegraphics[scale=0.5]{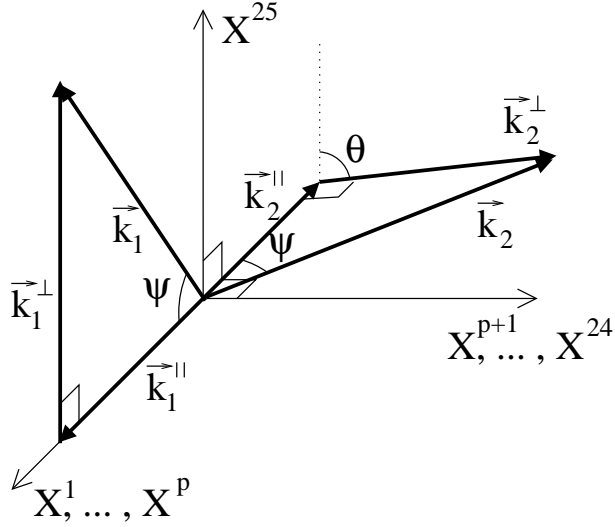}
\caption{This figure illustrates the kinematic variables: the spatial momenta of
the emitted strings $\vek_i$ with the angles $\theta ,\psi$.}
\label{fig:kin}
\end{figure}

To conclude our analysis, let us consider the asymptotic behavior of the $\xi=\xi_2$ amplitude \eq{bulk2ptfin} in different kinematic limits. We notice that the kinematic variables can be written as
\bea
 s &=& 4-2(\vek^{||})^2- (\vek^\perp)^2(1-\cos\theta) \nonumber \\
 t &=& 8-2(\vek^\perp)^2 \nonumber \\
 u &=& (\vek^\perp)^2(1-\cos\theta) \ ,
\eea
where $(\vek^\perp)^2 \equiv (\vek_1^\perp)^2 = (\vek_2^\perp)^2$ and $\theta$ is the angle between $\vek_1^\perp$ and $\vek_2^\perp$ (see figure~\ref{fig:kin}).
Before going to the explicit asymptotics, we make an important general observation:
the amplitude in \eq{bulk2ptfin} vanishes exponentially ( $\sim e^{-2\pi\om}$ or faster) in the high energy limit. Recall that one-point amplitude \cite{Lambert:2003zr} behaves as $\sim e^{-\pi\om}$. This result follows since the sums of the arguments of the two beta functions in \eq{bulk2ptfin} are bounded from above,
\bea
 \frac{t+u}{2}-2 &=& 2 -(\vek^\perp)^2 \frac{1+\cos\theta}{2} \le 2\nonumber \\
 \frac{s+u}{2}   &=& 2 - (\vek^{\pp})^2  \le  2 \ ,
\eea
which excludes the region where the beta functions would grow exponentially for $\om \to \infty$. They can have a power-like behavior in $\om$ if at least one of the sums is fixed for $\om \to \infty$, in which case the behavior $\sim e^{-2\pi\om}$ arises from the factor $1/\sinh2\pi\om$ that multiplies the beta functions. The same factor is responsible for the suppression of the one-point amplitude. In both cases the exponent
contains the total energy of the emitted strings. In particular
this suggests that in the brane decay the two-string emission channel will be as important as the one-string
channel. However, we would need to study decay into massive string states and the phase space factors
before drawing conclusions.  %Next we study some explicit examples which are in accord with these general expectations.

Let us consider the following special limits:
\begin{itemize}
 \item[I    ] $(\vek^{||})^2 \to \infty$ with $(\vek^\perp)^2$ fixed, corresponding to string production parallel to the decaying brane
 \item[II\,] $(\vek^{\perp})^2 \to \infty$ with $(\vek^{\pp})^2$ (and $\theta$) fixed, corresponding to string production perpendicular to the decaying brane
 \item[III] $(\vek^{\perp})^2 \to \infty$ with $(\vek^{\pp})^2$ and $(\vek^{\perp})^2(1+\cos\theta)$ (or equivalently $\vek_1^\perp+\vek_2^\perp$) fixed, corresponding to string production perpendicular to the decaying brane such that the backreaction to the brane remains finite.
\end{itemize}
Notice that all these limits are reasonable, since $\om^2 = \vek^2-4 \to \infty$ in each case validating the electrostatic approximation. Recall that any subleading terms of the electrostatic approximation were not included, and hence results for subleading asymptotics (which we shall anyhow write down) are not reliable. Also notice that there are plenty of poles in the physically meaningful region. We only state the leading overall behavior ``averaged over the poles'', so the subleading terms may be significant very near their poles.

The limit I means taking $s \to -\infty$ with $t$ and $u$ fixed.
For $u>2$ the factor in the square brackets of \eq{bulk2ptfin} approaches a constant since the latter, $s$ dependent term decays,
\be
  A_{2}  \simeq  -\frac{ i \pi e^{i\frac{\pi s}{2}}(2\pi\la)^{2 i\omega}}{2\sinh2\pi\omega}\, e^{i\frac{\pi u}{2}}\frac{\Gamma\left(\frac{t}{2}-1\right)\Gamma\left(\frac{u}{2}-1\right)}{\Gamma\left(\frac{t+u}{2}-2\right)} \ ,
\ee
while for $0<u<2$ the latter term dominates and we find
\be
  A_{2}  \simeq  \frac{ i \pi e^{i\frac{\pi s}{2}}(2\pi\la)^{2 i\omega}}{2\sinh2\pi\omega}\,  \Gamma\left(\frac{u}{2}-1\right) \frac{\sin[\pi(s+u)/2]}{\sin[\pi s/2]}\left(-\frac{s}{2}\right)^{1-\frac{u}{2}} \ .
\ee
Thus for any $u$ the absolute value of the amplitude vanishes exponentially $\sim e^{-2\pi\om}$, as expected. The amplitude has also a very rapidly varying phase $\sim e^{-i\pi \om^2}$. The subleading behavior which arises from the terms in the square brackets is likely to be modified by the nonleading terms of the electrostatic limit which we have neglected.

The limit II corresponds to  $s,t \to -\infty$ while $u \to +\infty$ such that $t+u \to -\infty$ and $s+u = 4 - 2(\vek^{||})^2$ remains fixed. Then the first term in the square brackets vanishes exponentially, and the contribution from the second term leads to
\be \label{eq:asform}
 A_{2}  \simeq \frac{ i \pi e^{i\frac{\pi s}{2}}(2\pi\la)^{2 i\omega}}{2\sinh2\pi\omega} \frac{\pi}{\Gamma\left(\frac{s+u}{2}\right)\sin \frac{\pi s}{2}}\left(-\frac{s}{2}\right)^{\frac{s+u}{2}-1} \ .
\ee
Hence the absolute value of the amplitude behaves as $\sim e^{-2\pi\om}$ also in this case.

The limit III is otherwise the same as the limit II except that $t+u = 8 -(\vek^\perp)^2(1+\cos\theta)$ is fixed. The asymptotics is given by
\bea
 A_{2}  &\simeq &\frac{ i \pi e^{i\frac{\pi s}{2}}(2\pi\la)^{2 i\omega}}{2\sinh2\pi\omega} \frac{\pi}{\Gamma\left(\frac{s+u}{2}\right)\sin \frac{\pi s}{2}}\left(-\frac{s}{2}\right)^{\frac{s+u}{2}-1}\ ; \quad \frac{1+\cos\theta}{2}>\frac{(\vek^{\pp})^2}{(\vek^\perp)^2} \\ \nonumber
 A_2 &\simeq& \frac{ i \pi e^{i\frac{\pi s}{2}}(2\pi\la)^{2 i\omega}}{2\sinh2\pi\omega}\frac{\pi e^{i\frac{\pi u}{2}}}{\Gamma\left(\frac{t+u}{2}-2\right)\sin \frac{\pi t}{2}}\left(-\frac{t}{2}\right)^{\frac{t+u}{2}-3}\ ; \quad \frac{1+\cos\theta}{2}<\frac{(\vek^{\pp})^2}{(\vek^\perp)^2} \ .
\eea
That is, in the former case we reproduce \eq{asform}. Again, the absolute value of the amplitude vanishes as $\sim e^{-2\pi\om}$.

Let us finally consider string production in an arbitrary direction. We take $\om \to \infty$ with the parametrization
\bea
 (\vek^{\pp})^2 &=& \vek^2 \cos^2\psi  =(4+\om^2)  \cos^2\psi  \nonumber \\
 (\vek^{\perp})^2 &=& \vek^2 \sin^2\psi = (4+\om^2) \sin^2\psi  \ ,
\eea
where $0 < \psi < \pi/2$ (such that the above cases I-III correspond to the endpoints, see Figure~\ref{fig:kin}). We find the leading one of the following asymptotics:
\bea
 A_2\! &\sim&\! (2\pi\la)^{2 i\omega} e^{-2\pi\om} e^{-i \pi \om^2 \cos^2\psi}  \left[\left(\frac{1+\cos\theta}{2}\right)^\frac{1+\cos\theta}{2}\left(\frac{1-\cos\theta}{2}\right)^\frac{1-\cos\theta}{2}\right]^{\om^2\sin^2\psi} \nonumber \\
 A_2\! &\sim&\! (2\pi\la)^{2 i\omega} e^{-2\pi\om} e^{-i \pi \om^2\left( \cos^2\psi+\frac{1-\cos\theta}{2}\sin^2\psi\right)}
\Bigg[\!\!\left(\frac{\frac{1\!-\!\cos\theta}{2}\sin^2\psi}{\cos^2\psi +\frac{1\!-\!\cos\theta}{2}\sin^2\psi}\!\right)^{\frac{1\!-\!\cos\theta}{2}\sin^2\psi} \nonumber \\
&&\times \left(\frac{\cos^2\psi}{\cos^2\psi +\frac{1\!-\!\cos\theta}{2}\sin^2\psi}\right)^{\cos^2\psi}\!\Bigg]^{\om^2} \ .
\eea
The amplitudes vanish very fast $\sim \exp(- C \om^2)$ due to the terms with the square brackets. It is easy to check that the coefficient $C$ is positive.

\bigskip
\noindent

{\bf \large Acknowledgments}

We thank Oren Bergman, Dennis D. Dietrich, Paolo Di Vecchia, Gilad Lifschytz, Matt Lippert, Sean Nowling, and Patta Yogendran for useful comments and discussions. N.J. has been supported in part by the Israel Science Foundation under grant no. 568/05 and in part at the Technion
by a fellowship from the Lady Davis Foundation. M.J. has been supported in part by the Marie Curie Excellence Grant under contract MEXT-CT-2004-013510 and in part by the Villum Kann Rasmussen foundation. E.K-V. has been supported in part by the Academy of Finland grant number
1127482. This work has also been supported in part by the EU 6th Framework Marie Curie Research and Training network ``UniverseNet''
(MRTN-CT-2006-035863).

%%%%%%%%%%%%%%%%%%%%%%%%%%%%%%%%%%%%%%%%%%%%%%%%%%%%%%%%%%%%%%%%%%%%%%%%%%%%%%%%%%%%%%%%%%%%%%%%%%%%%%
%%%%%%%%%%%%%%%%%%%%%%%%%%%%%%%%%%%%%%%%%%%%%%%%%%%%%%%%%%%%%%%%%%%%%%%%%%%%%%%%%%%%%%%%%%%%%%%%%%%%%%
%%%%%%%%%%%%%%%%%%%%%%%%%%%%%%%%%%%%%%%%%%%%%%%%%%%%%%%%%%%%%%%%%%%%%%%%%%%%%%%%%%%%%%%%%%%%%%%%%%%%%%
%%%%%%%%%%%%%%%%%%%%%%%%%%%%%%%%%%%%%%%%%%%%%%%%%%%%%%%%%%%%%%%%%%%%%%%%%%%%%%%%%%%%%%%%%%%%%%%%%%%%%%

\appendix

\section{Correlators and kinematics}\label{app:kin}

In this appendix we write the bulk two-point amplitude in terms of the Mandelstam variables often used in the literature.
Let us start by giving a heuristic argument to motivate the definition of the variables.
For the moment let us ignore the energy transfer from the D-brane to the strings which is possible due to the half S-brane deformation.
Consider the  $2\to 2$ scattering {\em tachyon} $(k_1)$ + {\em D-brane} $(p_1)
\rightarrow$ {\em tachyon} $(k_2)$ + {\em D-brane} $(p_2)$, with the associated momenta in
brackets. Breaking into parallel and perpendicular components,
\bea
k_1 &=& (k_1^{\pp},\vek_1^\perp) \nonumber \\
p_1 &=& (0,-\vek_1^\perp) \nonumber \\
k_2 &=& (k_2^{\pp},\vek_2^\perp) \nonumber \\
p_2 &=& (0,-\vek_2^\perp) \ .
\eea
On-shell conditions and momentum conservation in parallel directions give
\bea
k_1^2 &=& k_2^2 = 4 \nonumber \\
k_1^{\pp} &=& k_2^{\pp}\nonumber \\ |\vek_1^\perp| &=& |\vek_2^\perp| \ .
\eea
So we obtain the Mandelstam variables
\bea \label{eq:Manddef}
\tilde s &=& (k_1+p_1)^2 = (k_2+p_2 )^2 = (k_1^{\pp} )^2 = (k_2^{\pp} )^2 = \half s \nonumber \\
\tilde t &=& (k_1-k_2)^2 = (p_1-p_2)^2 = (\vek_1^\perp-\vek_2^\perp)^2 =
k_1^2+k_2^2-2k_1\cdot k_2 = 8 + 2t \nonumber \\
\tilde u &=& (k_1-p_2)^2 = (k_2-p_1)^2 = (k_1^{\pp} )^2 + (\vek_1^\perp+\vek_2^\perp)^2 =
16-\tilde t-3\tilde s \ ,
\eea
where the variables without tildes match with the conventions of \cite{Balasubramanian:2004fz}.\footnote{Notice that $k_2$ has opposite sign in \cite{Balasubramanian:2004fz}.}
The last line of \eq{Manddef} is consistent with the relation
\be
\tilde s+\tilde t+\tilde u = {\rm sum\ of\ squared\ masses} =  8 + (\vek_1^\perp)^2 + (\vek_2^\perp)^2 = 16 - 2\tilde s \ .
\ee

Consider then the bulk two-point tachyon amplitude
\bea\label{e1}
  A_{2} &= & \int dx^0 d\vec x_{\pp} e^{i(k_1+k_2)\cdot x_{\pp}}
 \Bigg\{ \int\prod_{a=1}^{2}\frac{d^2 w_a}{2\pi} \vev{\prod_{a=1}^2 e^{i k_a\cdot X'(w_a,\bar w_a)}}  \\
\mbox{} && + \sum_{N=1}^\infty\frac{(-z)^N}{N!}\int\prod_{i=1}^N\frac{dt_i}{2\pi}\prod_{a=1}^{2}\frac{d^2 w_a}{2\pi}
\vev{\prod_{i=1}^N e^{X'^0(t_i)}\prod_{a=1}^2 e^{i k_a\cdot X'(w_a,\bar w_a)}} \Bigg\} \ . \nonumber
\eea
The singular self-contractions will be dropped. The first term in the
series would give the amplitude for scattering from a stable D-brane. It contains
the correlator
\be
K_1 = \vev{\prod_{a=1}^2 e^{i k_a\cdot X'(w_a,\bar w_a)}} \ .
\ee
Working out the contractions gives
\bea
K_1 = \exp \Big\{\!\!\!\!\!\!\!\!\! && \half \sum_{a,b} \{k^{\pp}_a\cdot k^{\pp}_b [\ln |w_a-w_b|+\ln |1-w_a\bar w_b|] \nonumber \\
 && + \vek^{\perp}_a\cdot \vek^{\perp}_b [\ln |w_a-w_b|-\ln |1-w_a\bar w_b|]\} \Big\} \ . \nonumber
\eea

For the next term in (\ref{e1}) we need to evaluate the correlator
\be
 K_2 =  \vev{\prod_{i=1}^N e^{X'^0(t_i)}\prod_{a=1}^2 e^{i k_a\cdot X'(w_a,\bar w_a)}} \ .
\ee
We find
\be
K_2 = \prod_{i<j} |e^{it_i}-e^{it_j}|^2 \prod_{ia} |1-w_ae^{-it_i}|^{2\xi_a} \cdot K_1 \ ,
\ee
where $\xi_a = -i\om_a$. Let us naively generalize the above definitions of the Mandelstam variables to the case where the energy transfer is not zero by
\bea
   k_1\cdot k_2 &=& s \nonumber \\
   (k^{\pp}_1)^2 -(\vek^{\perp}_1)^2 &=& t-4 \ ,
\eea
where we crossed $s \leftrightarrow t$ to obtain a setting that corresponds to the production of two strings.
Then, setting $w_1=r,w_2=0$, we get for the amplitude
\bea\label{e2}
  A_{2} =  \delta (\vec{k}^{\pp}_1+\vec{k}^{\pp}_2) \int dx^0 e^{(\xi+\xi_2) x^0}\sum_{N=0}^{\infty}
  (-z)^N T(N) \ ,
\eea
where
\be
 T(N=0)=T(N=1)=\int drr~K_1(r)=\int drr~r^s (1-r^2)^{(t/2)-2}
\ee
and for $N\ge 2$,
\be
 T(N)= \int drr~r^s (1-r^2)^{(t/2)-2}\Big( \frac{1}{N!}\Big)\int \prod_{i=1}^N \frac{dt_i}{2\pi} \prod_{i<j} |e^{it_i}-e^{it_j}|^2
 \prod_i |1-re^{-it_i}|^{2\xi} \ .
\ee

\section{Schur polynomial method}\label{app:Schur}

Let us first
discuss some properties of the Schur polynomials.
They are symmetric polynomials of $N$ variables $x_1,\ldots x_N$, denoted by $s_\la(x_1,x_2,\ldots,x_N)$ where $\la=(\la_1,\la_2,\ldots,\la_N)$ is a partition of $|\la|=\la_1+\la_2+\cdots +\la_N$ with $\la_1 \ge \la_2 \ge \cdots \ge \la_N \ge 0$. The partition may be identified with a Young diagram where the $i$th row has $\la_i$ boxes. The polynomials may be defined by
\be
 s_\la(x_i) = \frac{\det\left[ x_i^{\la_j+N-j}\right]_{i,j=1,\ldots N}}{\det\left[ x_i^{N-j}\right]_{i,j=1,\ldots N}} \ ,
\ee
where the denominator is the Vandermonde determinant $\Delta(x_1,x_2,\ldots,x_N)$.
The definition may be extended to partitions having more than $N$ nonzero elements $\la_i$ (or to Young diagrams having more than $N$ rows) by setting these polynomials to zero. We denote by $\ell(\la)$ the index of the last nonzero element (or the height of the corresponding Young diagram).

The polynomials have the following well known identities \cite{Forrester}
\bea \label{prodf}
 \prod_{i=1}^{N_x}\prod_{j=1}^{N_y}(1+x_iy_j) &=& \sum_\la s_\la(x_i) s_{\la'}(y_i) \\
 \prod_{i=1}^{N_x}\prod_{j=1}^{N_y}\inv{1-x_iy_j} &=& \sum_\la s_\la(x_i) s_{\la}(y_i) \ ,
\eea
where
$\la'$ is the conjugate partition of $\la$, \ie, $\la'_i$ are the number of boxes in the columns of the Young diagram of $\la$. Notice that since $s_\la(x_i)$ vanishes if $\ell(\la)>N_x$, the sum over $\la$ is restricted to a finite set of partitions with $\ell(\la)\le N_x$ and $\la_1\le N_y$ in the first identity, as expected since the left hand side is a polynomial in $x_i$ and $y_j$.
One can also write down a generalized binomial theorem
\be\label{Schurid}
 \prod_{i=1}^N (1-x_i)^\xi = \sum_\la c_\la(\xi) s_\la(x_i) \ ,
\ee
where the coefficients $c_\la(\xi)$ are independent of $N$ and polynomials in $\xi$. They are well known and are related to (\ref{prodf}).
When $\xi$ is a positive integer we can set in (\ref{prodf}) $N_y=\xi$ and $y_1=-1=y_2=\cdots =y_\xi$ whence we see that the coefficient $c_\la(\xi)$ equals $s_{\la'}(y_i=-1)$. The result may be expressed in various forms:
\bea
 c_\la(\xi) &=& (-1)^{|\la|} s_{\la'}(y_i=1) =  (-1)^{|\la|} \prod_{1\le i<j \le \xi }\frac{\la'_i-i-\la'_j+j}{j-i}\nonumber\\
 &=&  (-1)^{|\la|}\frac{\Delta\left(\la'_\xi+1,\la'_{\xi-1}+2,\ldots,\la'_1+\xi\right)}{(\xi-1)!\cdots 1!}
\nonumber\\
 &=& (-1)^{|\la|} \prod_{m \in \la} \inv{h_m} (\xi +i(m)-j(m)) \ ,
\eea
where the last form gives the number of terms in the Schur polynomial in terms of the ``hook length'' $h_m$  (it also equals the number of the semistandard Young tableaux of the form $\la'$ and the dimension of the representation of $SU(\xi)$ characterized by $\la'$). Here $i(m)$ ($j(m)$) is the number of row (column) of the box $m$ in $\la$.
Since $c_\la$ are indeed known to be polynomials in $\xi$, the result in the last form readily extends to all complex values of $\xi$. We may thus write the generalized binomial theorem as
\be\label{Schurid2}
 \prod_{i=1}^N (1-r x_i)^\xi = \sum_\la (-r)^{|\la|}
 \prod_{m \in \la} \inv{h_m}
 (\xi +i(m)-j(m))\ s_\la(x_i) \ ,
\ee
where we rescaled all variables by $r$.

Let us then concentrate on the two-point function $Z_2$ of \eq{Z2def}. By using the formulae (\ref{Schurid2})
we may write the ``cross-term'' in the integrand as
\be
  \prod_{i=1}^N |1-r e^{-it_i}|^{2\xi} = \prod_{i=1}^N (1-r e^{it_i})^{\xi}(1-r e^{-it_i})^{\xi} = \sum_{\la,\bar \la} (-r)^{|\la|+|\bar \la|}c_\la(\xi) c_{\bar \la} (\xi) s_\la(x_i) s_{\bar \la}(x_i^*) \ ,
\ee
where $x_i=e^{it_i}$. As the Schur polynomials are orthogonal with respect to the weight function of $Z_2$ we immediately get
\be \label{eq:Z2intN}
 Z_2(r,\xi;N) = \sum_{\la,\ell(\la)\le N} r^{2|\la|} \left|c_\la(\xi)\right|^2 =\sum_{\la,\ell(\la)\le N} r^{2|\la|} \prod_{m \in \la} \frac{1}{h_m^2} (\xi +i(m)-j(m))^2\ ,
\ee
where the constraint $\ell(\la)\le N$
appears since the polynomials $s_\la(x_i)$ vanish for $\ell(\la)> N$. If $\xi$ is a positive integer the sum over $\lambda$ is finite because the coefficients $c_\la(\xi)$ vanish for $\la_1 > \xi$. Moreover, if $\xi$ is a negative integer and $N \ge |\xi|$ the sum can be done explicitly as a special case of the general result found in \cite{Constable:2003rc}:
\be
 Z_2(r,\xi;N) = (1-r^2)^{-\xi^2} \ ;  \qquad \xi=-1,-2,\ldots,-N \ .
\ee

%%%%%%%%%%%%%%%%%%%%%%%%%%%%%%%%%%%%%%%%%%%%%%%%%%%%%%%%%%%%%%%%%%%%%%%%%%%%%%%%%%%%%%%%%%%%%%%%%%%%%%%%%%%%%%%%%%%%%%%%%%%%%%%%%%%%%%%%%%%%%%%%%%%%%%%%%%%%%%%%%%%%%%%%%%%%%%%%%%%%%%%%%%%%%%%%%%%%%%%%%%%%%%%%%%%%%%%%%%%%%%%%%%%%%%%%%%%%%%%%%%%%%%%%%%%%%%%%%%%%%%%%%%%%%%%%%%%%%%%%%%%%%%%%%%%%%%%%%%%%%%%%%%%%%%%%%%%%%%%%%%%%%%%%%%%%%%%%%%%%%%%%%%%%%%%%%%%%%%%%%%%%%%%%%%%%%%%%%%%%%%%%%%%%%%%%%%%%%%%%%%%%%%%%%%%%%%%%%%%%%%%%%%%%%%%%%%%%%%%%%%%%%%%%%%%%%%%%%%%%%%%%%%%%%%%%

\section{Solutions to the potential problems} \label{app:genmet}

We are interested in solving potential problems in classical electrostatics where the conductors are pieces of the arc of the unit circle.
The general potential problem has as conductors $\hat n$ separate arcs of the unit circle. We find it useful to map the disk onto the upper half plane by the conformal map
\be \label{eq:confmap}
 w \mapsto z = i \beta \frac{1-w}{1+w} \equiv q(w)
\ee
with the understanding that we work in the compactified complex plane with the $\infty$-point included. Here $\beta$ is a real parameter.

The analytic properties of the complex electric field allow us to write down the solution in a special form \cite{Forrester}. Let us assume for a moment that we have two conductors, $\hat n=2$, which lie at $[a,b]$ and at $[c,d]$. We can then define the following functions
\bea
 g(z) &=& \sqrt{z-a} \sqrt{z-b}\sqrt{z-c}\sqrt{z-d} \nn\\
 h(z) &=& \sqrt{z-a} \sqrt{b-z}\sqrt{c-z}\sqrt{d-z} \ ,
\eea
where the principal branch of the square root function is used.
Then $g(z)$ is analytic everywhere except on the conductors where it has branch cuts, whereas $h(z)$ is analytic over the conductors, but has branch cuts in the gap regions.

Generalization to $\hat n$ conductors is found by adding more terms to $g(z)$ and $h(z)$. We choose the branches such that
\be \label{eq:hgrel}
 h(z) = \mp i g(z) \ ,
\ee
where the minus sign holds in the upper half plane and the plus sign holds in the lower half plane.
It is then possible to show that, when exposed to an appropriate external (conjugate) electric field $E_{\rm ext}(z)$, a solution for the (conjugate) electric field due to the conductors reads
\be
 E_c(z) = \inv{\pi} g(z) \int_\mathrm{cond} dt \frac{\re E_{\rm ext}(t)}{(z-t)h(t)} \ ,
\ee
where the integration is over the conductors. Notice that the solution is bounded at the conductor endpoints.
It is well defined everywhere except on the conductors, where it has discontinuities. The charge density\footnote{For brevity, we omit the subscript $0$ which was used in the main text.} is proportional to the discontinuity and becomes
\be
 \rho(z) =  \inv{\pi^2}h(z) \int_\mathrm{cond} dt \inv{h(t)} \frac{\re E_{\rm ext}(z)-\re E_{\rm ext}(t)}{z-t} \ .
\ee

The $2\hat n$ endpoints of the conductors are not arbitrary parameters but are subject to constraints.
In particular, these parameters (the constants $a$, $b$, $c$, and $d$ in the $\hat n=2$ case) are partially fixed by the asymptotics of the electric field
\be
 E_c(z) \sim \frac{N}{z}
\ee
for $z \to \infty$ where $N$ is the total electric charge on the conductors. This gives
\bea \label{eq:constraints}
0 &=& \int_\mathrm{cond} dt \frac{t^k\re E_{\rm ext}(t)}{h(t)} \ ,\qquad k=0, \ldots ,\hat n-1 \nn\\
\pi N &=& \int_\mathrm{cond} dt \frac{t^{\hat n}\ \re E_{\rm ext}(t)}{h(t)} \ .
\eea
There are $\hat n-1$ parameters left free, which are identified as the potential differences between the conducting planes. We will set these to zero and thus require
\be
 \re \int_{\mC_k} dz E(z) = 0\ , \qquad k=2, \ldots ,\hat n  \ ,
\ee
where
\be
 E(z) = E_c(z)+ E_{\rm ext}(z)
\ee
is the total electric field and the curve $\mC_k$ connects the first conductor with the $k$th one without intersecting the other conductors, for example.

For the case of interest to us, the external field is due to $n$ point like charges,
\be
 E_{\rm ext}(z) = \sum_{a=1}^n \xi_a \inv{z-z_a} \ .
\ee
For this configuration the integrals above can be calculated exactly by using contour integration methods. We find
\bea \label{eq:Esol}
 E_c(z) & = & \frac{g(z)}{2}\sum_a\xi_a \left[\inv{g(z_a)(z-z_a)}+\inv{g(z_a^*)(z-z_a^*)}\right]-\sum_a\xi_a \frac{z-x_a}{(z-z_a)(z-z_a^*)} \nn\\
 E(z)   & = &  \frac{g(z)}{2}\sum_a\xi_a \left[\inv{g(z_a)(z-z_a)}+\inv{g(z_a^*)(z-z_a^*)}\right]+\sum_a\xi_a \frac{iy_a}{(z-z_a)(z-z_a^*)} \nn\\
\rho(z) & = & -\frac{h(z)}{2\pi}\sum_a\xi_a \left[\inv{g(z_a)(z-z_a)}+\inv{g(z_a^*)(z-z_a^*)}\right] \ ,
\eea
where $z_a=x_a+iy_a$. The asymptotic conditions imply
\bea \label{eq:constraints2}
 0             & = & \sum_a\xi_a \left[\frac{z_a^k}{g(z_a)}+\frac{\left(z_a^*\right)^k}{g(z_a^*)}\right], \qquad k=0, \ldots, \hat n-1 \nn\\
 N+\sum_a\xi_a & = & \inv{2}\sum_a\xi_a \left[\frac{z_a^{\hat n}}{g(z_a)}+\frac{(z_a^*)^{\hat n}}{g(z_a^*)}\right] \ .
\eea

The potential and the total energy of the system are of special interest to us. We can integrate the complex electric field in \eq{Esol} to give
\bea \label{eq:Vsol}
 V_c(z) & = &  V(z) + \sum_a \xi_a \log(z-z_a) \nn\\
 V(z)   & = & \inv{2}\sum_a\xi_a\left[-\log(z-z_a)+\log(z-z_a^*)\right] \nn\\
        &   & +\lim_{R \to \infty} \Bigg\{\inv{2}\int_z^R dt g(t)\sum_a\xi_a  \left[\inv{g(z_a)(t-z_a)}+\inv{g(z_a^*)(t-z_a^*)}\right]\nn\\
        &   & - \left(N+\sum_a\xi_a\right)\log R\Bigg\} \ .
\eea
Notice that this solution has the asymptotics required by  our definition \eq{Vdef} (or more precisely by its natural generalization to the half plane)
\beq
  V(z)  =  - \left(N+\sum_a\xi_a\right)\log z + \morder{\frac{1}{z}} \ ,
\eeq
so that the constant term of the expansion in $1/z$ is zero.

Let us then write down explicitly the solutions for the disk, \ie, apply the transformation $q(w)$ of \eq{confmap} (with, say $\beta=1$). To do so, we should start with the configuration where charges $\xi_1,\ldots, \xi_n$ are located at arbitrary points $z_1 = q(w_1),\ldots , z_n= q(w_n)$, and an additional charge $\xi_{n+1}=-N - \sum_a \xi_a$ is located at $z=q(\infty)$. Let us denote the corresponding electric field and potential, as given by eqs.~\eq{Esol} and~\eq{Vsol}, by $\tilde E$ and $\tilde V$, respectively.

The electric field on the disk may be then written as
\beqa \label{eq:Emap}
 E(w)   &=& \tilde E(q(w)) \nn\\
 E_c(w) &=& \tilde E_c(q(w)) - \frac{N}{q(w)-q(\infty)}
\eeqa
and integration gives the potential on the disk
\beqa \label{eq:VVcdef}
 V(w)   &=& \tilde V(q(w)) - \lim_{R \to \infty}\left[\int_{q(R)}^\infty dz \tilde E(z) +\left(N+\sum_a\xi_a\right)\log R\right] \nn\\
 V_c(w) &=& \tilde V_c(q(w)) + N \log \left[q(w)-q(\infty)\right]  \nn\\
        &&- \lim_{R \to \infty}\left[\int_{q(\infty)}^R dz \tilde E(z) -N \log R\right] - N\log \left.\frac{d}{dt}\, q\left(1/t \right)\right|_{t=0}  \ .
\eeqa
Here the integrals are independent of $w$. They were added to achieve the correct asymptotics,
\beq
 V(w) =  - \left(N+\sum_a\xi_a\right)\log w + \morder{\frac{1}{w}} \ ; \qquad V_c(w) =  - N \log w + \morder{\frac{1}{w}} \ .
\eeq

Finally, the potential on the conductors is found as
\beq \label{eq:Udef}
 U_0 =\re\, V(w_0)\ ,
\eeq
where $w_0$ is any point on the conductors, and the ``interaction'' energy is given by
\beq \label{eq:Ecxidef}
 \mE_{c-\xi} = \sum_a \xi_a \re \, V_c(w_a) \ .
\eeq
The total energy then reads (see \eq{Etotres} above)
\beq
 \mE_{\rm tot} = \frac{N}{2}U_0 + \inv{2}\mE_{c-\xi} + \mE_{\xi} \ ,
\eeq
where
\be
 \mE_\xi = -\sum_{a < b} \xi_a\xi_b \log(w_a-w_b)
\ee
is the energy linked to the self-interactions of the external charges. This trivial term was not discussed in the main text
since its contribution was embedded in the explicit $w_1,w_2$ dependence of \eq{Abardef} that was treated separately.

\end{document}